\documentclass[11pt]{article}

\usepackage{graphicx}
\usepackage{amsmath,amssymb,array}
    \RequirePackage{fancyvrb}
    \RequirePackage{alltt}
    \DefineVerbatimEnvironment{example}{Verbatim}{}
    \renewenvironment{example*}{\begin{alltt}}{\end{alltt}}
\usepackage[utf8]{inputenc}
\usepackage{color}

\usepackage[top=2cm, bottom=2cm, left=2cm, right=2cm]{geometry}

\title{\textit{multiColl} package and other packages to detect multicollinearity in \textbf{R}}

\author{Román Salmerón \\
              Department of Quantitative Methods for the Economy and Business \\
              University of Granada (Spain) \\
              Tel.: +34-958248791\\
              email: romansg@ugr.es \\
           Catalina García  \\
              Department of Quantitative Methods for the Economy and Business \\
              University of Granada (Spain) \\
              Tel.: +34-958248790\\
              email: cbgarcia@ugr.es \\
           José García \\
              Department of Economics and Business \\
              University of Almería (Spain) \\
              Tel.: +34-958248790\\
              email: jgarcia@ual.es
}

\begin{document}

\maketitle

\begin{abstract}
    This work presents a guide for the use of some of the functions of the \textbf{multiColl} \cite{multiCollR} package in  \textbf{R} \cite{R} for the detection of near-multicollinearity.  The main contribution, in comparison to other existing packages in \textbf{R} or other econometric software, is the treatment of qualitative independent variables and the intercept in the simple/multiple linear regression model. The main goal of this paper is to show the advantages of the \textbf{multiColl} package in \textbf{R}, comparing its results with the results obtained by other existing packages in \textbf{R} for the treatment of multicollinearity.
\end{abstract}


\section{Introduction}

Given a model with $n$ observations and $k$ independent variables is specified as follows:
\begin{equation}
\label{model0}
\mathbf{y}_{n \times 1} = \mathbf{X}_{n \times k} \cdot \boldsymbol{\beta}_{k \times 1} + \mathbf{u}_{n \times 1},
\end{equation}
where the first column of $\mathbf{X}$ is composed of ones representing the intercept and $\mathbf{u}$ represents the random disturbance assumed to be centered and spherica. That is, $E[\mathbf{u}_{n \times 1}] = \mathbf{0}_{n \times 1}$ and $var(\mathbf{u}_{n \times 1}) = \sigma^{2} \cdot \mathbf{I}_{n \times n}$, where $\mathbf{0}$ is a vector of zeros, $\sigma^{2}$ is the variance of the random disturbance, and $\mathbf{I}$ is the identity matrix. 

The aim is to estimate the coefficients $\boldsymbol{\beta}$ of the independent variables and, from these values, establish the direction of relations (according to the signs) and quantify relations (with values).
The ordinary least squares (OLS) approach is the most frequently applied methodology for obtaining the estimates of coefficients where it is assumed that the variables in matrix $\mathbf{X}$ are independent. Otherwise, it is said that the model is characterized by near-multicollinearity.

The structure of the paper is as follows: Section \ref{function.detection} presents how to use the \textbf{multiColl} \cite{multiCollR} package to calculate the measures most commonly applied in the scientific literature to detect the presence of multicollinearity by paying special consideration to qualitative explanatory variables which are usually ignored in the existing software.
The different functions are compared with the functions included on the  \textbf{car} \cite{car2018}, \textbf{mctest} \cite{mctest2018}, \textbf{mcvis} \cite{mcvis} and \textbf{rms} \cite{rms} packages in \textbf{R} which are also applied to obtain similar measures.
Section \ref{function.slm} presents the particular case of the simple linear regression where non-essential multicollinearity (the relationship between the intercept and the rest of the independent variables) can exist although  it is ignored in econometric software such as  \textbf{GRETL} \cite{GRETL} (see \cite{Salmeron2019b} for more details) or the \textbf{car} \cite{car2018}, \textbf{mctest} \cite{mctest2018}, \textbf{mcvis} \cite{mcvis} and \textbf{rms} \cite{rms}  packages in \textbf{R}  as is shown in this paper.
Finally,  Section \ref{conclusiones} summarizes the main contributions of this paper.

\section{Multicollinearity detection}
    \label{function.detection}

\cite{version_arxiv} analyzes the calculation by \textbf{multiColl} \cite{multiCollR} package of the correlation matrix of a model's independent variables and its determinant, the variance inflation factors, the condition number (with and without the intercept) and the Stewart index, among others.
In this section, these functions are compared to similar functions existing in other  packages in \textbf{R}, paying special attention to the treatment of quantitative variables and the intercept.
More concretely, are analyzed the functions existing in the \textbf{car} \cite{car2018}, \textbf{mctest} \cite{mctest2018}, \textbf{mcvis} \cite{mcvis}, \textbf{rms} \cite{rms} and \textbf{perturb} \cite{perturb2012}  packages in \textbf{R} to the Henri Theil's \cite{Theil:1971} textile consumption data (see  Table \ref{theil}) to analyze  how these packages treat the dummy variable \textit{twentys} and if these packages are able to detect the non-essential multicollinearity existing due to the slight variability of variable \textit{income}.


\begin{table}
    \centering
    \caption{Henri Theil's textile consumption data} \label{theil}
    \begin{tabular}{ccccc}
        \hline\noalign{\smallskip} 
        Year & Consumption & Income & Relprice & Twenties \\
        \noalign{\smallskip}\hline\noalign{\smallskip} 
        1923 & 99.2 & 96.7 & 101.0 & 1 \\
        1924 & 99.0 & 98.1 & 100.1 & 1 \\
        1925 & 100.0 & 100.0 & 100.0 & 1 \\
        1926 & 111.6 & 104.9 & 90.6 & 1 \\
        1927 & 122.2 & 104.9 & 86.5 & 1 \\
        1928 & 117.6 & 109.5 & 89.7 & 1 \\
        1929 & 121.1 & 110.8 & 90.6 & 1 \\
        1930 & 136.0 & 112.3 & 82.8 & 0 \\
        1931 & 154.2 & 109.3 & 70.1 & 0 \\
        1932 & 153.6 & 105.3 & 65.4 & 0 \\
        1933 & 158.5 & 101.7 & 61.3 & 0 \\
        1934 & 140.6 & 95.4 & 62.5 & 0 \\
        1935 & 136.2 & 96.4 & 63.6 & 0 \\
        1936 & 168.0 & 97.6 & 52.6 & 0 \\
        1937 & 154.3 & 102.4 & 59.7 & 0 \\
        1938 & 149.0 & 101.6 & 59.5 & 0 \\
        1939 & 165.5 & 103.8 & 61.3 & 0 \\
        \noalign{\smallskip}\hline 
    \end{tabular}
\end{table}


Lastly, we should indicate that the help instructions for the \textbf{multiColl} \cite{multiCollR} package can be found in:\\
$https://cran.r-project.org/web/packages/multiColl/multiColl.pdf.$

\subsection{The \textbf{car} and \textbf{rms} packages}

By using the  \textit{vif} command of the \textbf{car} \cite{car2018} package it is possible to obtain the VIFs without any consideration:
\begin{example}
>     reg.theil = lm(consume~income+relprice+twentys)
>     vif(reg.theil)
  income relprice  twentys
1.062760 6.007181 5.866333
\end{example}
However, the VIF associated to the variable \textit{twentys} is obtained from the coefficient of determination of a regression whose dependent variables is dichotomous. As is well known, models are not linear in this kind of fit, and, for this reason, it is not appropriate to use the coefficient of determination in this case. Identical results are obtained by using the \textit{vif} command of the \textbf{rms} \cite{rms} package. In addition, none of these packages indicate in their help instructions that the VIF is only suitable for explanatory quantitative variables.

Note that, to the best of our knowledge, the Stewart index is not calculated in any other  package in \textbf{R}. It only can be obtained manipulating the \textit{vif} command of the \textbf{rms} \cite{rms} package. This manipulation consists in introducing the intercept as an independent variable within the matrix $\mathbf{X}$  and indicating that the model does not have an intercept with the \textit{lm} command:
\begin{example}
> reg.0 = lm(theil.y~theil.X+0)
> vif(reg.0)
        theil.X   theil.Xincome theil.Xrelprice  theil.Xtwentys
     427.445968      427.228985      136.668316        9.972766
\end{example}
As shown by \cite{Salmeron_2020}, after this manipulation the model is considered to be non centered and, consequently, the  \textit{vif} command will be calculating the Stewart index instead of VIF. Thus, in this situation the researcher wrongly considers that is calculating the VIF when, in fact, the Stewart index was obtained. Apart from this limitation, these packages do not handle non-quantitative variables properly either.

\subsection{The \textbf{mctest} package}

The \textit{mctest} command  from the \textbf{mctest} \cite{mctest2018} package allows us to obtain a great number of measures to detect multicollinearity both from an individual and a joint point of view:
\begin{example}
> mctest(reg.theil)

Call:
omcdiag(mod = mod, Inter = TRUE, detr = detr, red = red, conf = conf,
    theil = theil, cn = cn)

Overall Multicollinearity Diagnostics

                       MC Results detection
Determinant |X'X|:         0.1650         0
Farrar Chi-Square:        25.5246         1
Red Indicator:             0.5371         1
Sum of Lambda Inverse:    12.9363         0
Theil's Method:           -0.1837         0
Condition Number:         53.3967         1

1 --> COLLINEARITY is detected by the test
0 --> COLLINEARITY is not detected by the test
\end{example}

and from the individual point of view:
\begin{example}
> mctest(reg.theil, type="i", corr=TRUE)

Call:
imcdiag(mod = mod, method = method, corr = TRUE, vif = vif, tol = tol,
    conf = conf, cvif = cvif, ind1 = ind1, ind2 = ind2, leamer = leamer,
    all = all)

All Individual Multicollinearity Diagnostics Result

            VIF    TOL      Wi      Fi Leamer    CVIF Klein   IND1
income   1.0628 0.9409  0.4393  0.9414 0.9700 -0.0718     0 0.1344
relprice 6.0072 0.1665 35.0503 75.1077 0.4080 -0.4060     0 0.0238
twentys  5.8663 0.1705 34.0643 72.9950 0.4129 -0.3964     0 0.0244
           IND2
income   0.1029
relprice 1.4520
twentys  1.4451

1 --> COLLINEARITY is detected by the test
0 --> COLLINEARITY is not detected by the test

twentys , coefficient(s) are non-significant may be due to multicollinearity

R-square of y on all x: 0.9529

* use method argument to check which regressors may be the reason of collinearity
===================================

Correlation Matrix
             income  relprice    twentys
income   1.00000000 0.1788467 0.09351197
relprice 0.17884669 1.0000000 0.90809254
twentys  0.09351197 0.9080925 1.00000000

====================NOTE===================

relprice and twentys may be collinear as |0.908093|>=0.7
\end{example}
Similar to packages previously analyzed, it is not possible to avoid the dichotomous variable and it is considered to calculate the matrix of simple correlations and its determinant.
It even indicates that \textit{relprice} and \textit{twentys} may be collinear as their coefficient of simple correlation is higher than 0.7. Thus, it is not only calculating a  correlation coefficient between a quantitative variable and a qualitative one but it is also using a threshold (0.7) very far away from the one proposed by  \cite{Garcia:2018} ($\sqrt{0.9}$). In addition, the VIF is also calculated for the dichotomous variable.

Finally, note that with this command the variable \emph{twentys} is designated as the one which is responsible for the existing multicollinearity.  However, from the results previously presented the multicollinearity is generated by the relation between the intercept and \textit{income}.

\subsection{The \textbf{mcvis} package}

Contradictory results are obtained by using the \textit{mcvis} command from the \textbf{mcvis} \cite{mcvis} package. When original data (without transformations) are used, it is not able to detect the existing non-essential multicollinearity, apart from obtaining numerous warnings:
\begin{example}
> mcvis1 = mcvis(theil.X, standardise_method="none")
There were 50 or more warnings (use warnings() to see the first 50)
>   mcvis1
          income relprice twentys
tau4 0.02   0.26     0.36    0.36
\end{example}
According to the authors ``larger values indicate the greater contribution of the variable explaining the observed severity of multicollinearity''. Thus, it will be designating (similar to what occurred with the \textbf{mctest} package) \textit{relprice} and \textit{twentys} as the variables responsible for multicollinearity, leaving the intercept in the last place.

These results should not surprise us if we take into account that the values are related to the VIFs (which is unable to detect the non-essential multicollinearity) and the eigenvalues of matrix $\mathbf{X}^{t} \mathbf{X}$.

If data are transformed by the \textit{Euclidean} method, centered by mean and divided by Euclidean length, the intercept is eliminated of the analysis and, as consequence, the non-essential multicollinearity is also ignored. In any case, the results designate the variable \textit{income} as the one responsible for the multicollinearity although the reason will not be clear:
\begin{example}
>   mcvis2 = mcvis(theil.X[,-1], standardise_method="euclidean")
>   mcvis2
     income relprice twentys
tau3   0.89     0.08    0.04
\end{example}

If data are transformed with the default method, \textit{studentise}, centred by mean and divided by standard deviation, the intercept is also eliminated of the analysis and the variables \textit{relprice} and \textit{twentys} are designated as the ones responsible for the multicollinearity:
\begin{example}
>   mcvis3 = mcvis(theil.X[,-1])
>   mcvis3
     income relprice twentys
tau3   0.06     0.39    0.55
\end{example}
Although the authors,  \cite{mcvis}, acknowledge that ``there are different views on what centering technique is most appropriate in regression'' they are ignoring the problem stating that ``the role of scaling is not the focus of our work as our framework does not rely on any specific scaling method''.
However, it is necessary to take into account that the eigenvalues of matrix $\mathbf{X}^{t} \mathbf{X}$ change depending on the applied transformation (see, for example,  \cite{Salmeron__2018} for more details) while the VIF is invariant to these transformations (see  \cite{Garcia__2016}). For these reason, the results vary depending on the treatment given to the data.

The transformation of the variables is a common practice when multicollinearity is being treated and, for this reason, the \textit{mcvis} package should provide robust results for the different transformations or, at least, should clarify in which situations it needs to be used.

\subsection{ The \textbf{perturb} package}

The use of \textit{colldiag} command from the \textbf{perturb} \cite{perturb2012} package allows us to detect the relation between the intercept and the variable \textit{income} providing results that are complementary to the ones obtained by the \textbf{multiColl} \cite{multiCollR} package:
\begin{example}
> colldiag(reg.theil)
Condition
Index	Variance Decomposition Proportions
          intercept income relprice twentys
1   1.000 0.000     0.000  0.001    0.005
2   2.758 0.001     0.001  0.000    0.168
3  25.967 0.053     0.055  0.999    0.826
4  53.397 0.946     0.944  0.000    0.001
\end{example}
Thus, from the different commands existing in the analyzed  packages in \textbf{R} whose goal is to detect multicollinearity, this is the only one that allows detection of the existence of non-essential multicollinearity in the model.

\section{A special case, the simple linear regression model:  \textit{SLM} function}
    \label{function.slm}

The simple linear model (model (\ref{model0}) with $k=2$) is a particular case where it being systematically ignored in many different statistical software packages to determine if the near-multicollinearity is problematic. 
In this case, the condition number (CN), Stewart index and coefficient of variation (CV) can be useful to detect if the near-multicollinearity is problematic.
In the case of a dummy independent variable, the proportion of ones could be a good approximation to measure the relationship of this variable with the intercept, see Appendix \ref{ap}, replacing the CV.

In \cite{version_arxiv} the following models are analyzed: 
$$consumption = f(income), \quad consumption = f(relprice), \quad consumption = f(twentys).$$

In the first model, near non-essential multicollinearity is problematic, unlike in the second and third models. 

Then, this subsection applies the functions existing in the  \textbf{car} \cite{car2018}, \textbf{mctest} \cite{mctest2018}, \textbf{mcvis} \cite{mcvis}, \textbf{rms} \cite{rms} and \textbf{perturb} \cite{perturb2012} packages in \textbf{R} to the simple linear model $consumption = f(income)$ commented above.

\subsection{ The \textbf{car} and \textbf{mctest} package}

When the \textit{vif} command of the \textbf{car} \cite{car2018} package is applied, this eliminates the possibility that multicollinearity exists in this kind of model, since the following error is obtained:
\begin{example}
> reg.mls = lm(theil.y~theil.X[,2])
>   vif(reg.mls)
Error in vif.default(reg.mls) : model contains fewer than 2 terms
\end{example}

A similar result is obtained when the  \textit{mctest} command of  \textbf{mctest} \cite{mctest2018} package is applied:
\begin{example}
> mctest(reg.mls)
Error in if (ncol(x) < 2) stop("X matrix must contain more than one
    variable") : argument is of length zero

> mctest(reg.mls, type="i", corr=TRUE)
Error in if (ncol(x) < 2) stop("X matrix must contain more than one
    variable") : argument is of length zero
\end{example}

\subsection{ The \textbf{mcvis} package}

The problem of data transformation appears again when the \textit{mcvis} command of the \textbf{mcvis} \cite{mcvis} package is applied, since when the intercept is transformed a column of zeros is obtained. If the transformation is not applied, the command provides messages with numerous warnings and a result that could be interpreted as an indication of the existence of problematic near-multicollinearity:
\begin{example}
> mcvis = mcvis(theil.X[,1:2])
Error in svd(crossprodX1) : infinite or missing values in 'x'

> mcvis = mcvis(theil.X[,1:2], standardise_method="euclidean")
Error in svd(crossprodZ1) : infinite or missing values in 'x'

> mcvis = mcvis(theil.X[,1:2], standardise_method="none")
There were 50 or more warnings (use warnings() to see the first 50)
> mcvis
         income
tau2 0.5    0.5
\end{example}

However, the same result is obtained for the model $consumption = f(relprice)$ where it was previously established that the existence of multicollinearity is not worrying:
\begin{example}
> mcvis = mcvis(theil.X[,c(1,3)], standardise_method="none")
There were 50 or more warnings (use warnings() to see the first 50)
> mcvis
         relprice
tau2 0.5      0.5
\end{example}
This fact leads us to consider that this measure is not adequate for detecting multicollinearity in the simple linear model.

\subsection{ The \textbf{rms} package}

The following results are obtained by using the \textit{vif} command of the \textbf{rms} \cite{rms} package:
\begin{example}
>   reg.mls = lm(theil.y~theil.X[,2])
>   vif(reg.mls)
theil.X[, 2]
           1

>   reg.mls = lm(theil.y~theil.X[,1:2]+0)
>   vif(reg.mls)
      theil.X[, 1:2] theil.X[, 1:2]income
            401.9994             401.9994
\end{example}
In this case, it is possible to calculate the VIF being equal to 1 (which is in line with the results, previously commented, obtained by \cite{Salmeron2019b}).
In addition, it is possible to calculate the VIF if the intercept is included in the design matrix and it is indicated that the model does not have an intercept. However, as previously commented, these results are not the VIFs but the Stewart indices. Thus, it is important to take special care when interpreting this information due to the fact, for example, that it will not be appropriate to use the thresholds traditionally considered for the VIF.

\subsection{The \textbf{perturb} package}

Finally, the  \textit{colldiag} command of  the \textbf{perturb} \cite{perturb2012} package allows us to calculate the condition index and variance decomposition proportions without any considerations:
\begin{example}
> colldiag(reg.mls)
Index	Variance Decomposition Proportions
          intercept X
1   1.000 0.001     0.001
2  40.075 0.999     0.999
\end{example}
Thus, this command is the only one, except for the \textit{SLM} proposed by the \textbf{multiColl} \cite{multiCollR} package, that allows us to establish whether the degree of near-multicollinearity existing in the simple linear model is worrying.

\section{Conclusions}
    \label{conclusiones}

This paper presents some of the functions of the \textbf{multiColl} \cite{multiCollR} package to detect multicollinearity and compared these functions with similar functions existing in other packages in \textbf{R}.
The main contributions of this paper can be summarized as follows:
\begin{itemize}
    \item It has been clarified that the matrix of simple linear correlations, its determinant, the variance inflation factors and the Stewart index are not adequate when the model contains non-quantitative variables. In this case, unlike other packages existing in \textbf{R} to detect multicollinearity which have been analyzed in this paper, the \textbf{multiColl} \cite{multiCollR} package allows us to obviate this kind of variable.
    \item The superiority of the \textbf{multiColl} \cite{multiCollR} package has been shown in comparison to other packages existing in \textbf{R} (except for the \textit{colldiag} command of the \textbf{perturb} \cite{perturb2012} package) to detect the degree of multicollinearity existing in a simple linear regression. The main conclusion is that it is adequate to use the condition number with intercept, the Stewart index and the coefficient of variation. These measures have been calculated with the  \textit{SLM} command of the  \textbf{multiColl} \cite{multiCollR} package.
    \item Finally, to the best of our knowledge, none of the  packages in \textbf{R} allow us to calculate the Stewart index, with the only possibility being to manipulate the \textit{vif} command  of  the \textbf{rms} \cite{rms} package. This fact could be motivated by this measure having been erroneously identified with the variance inflation factor (see  \cite{Salmeron2019b} for more details). However, this manipulation can be dangerous in the hands of non-expert researchers who could consider that they have obtained the VIF (as indicated in the help section of this package) when in fact they had calculated the Stewart index.
\end{itemize}


%
\section*{Funding}

This work has been supported by project PP2019-EI-02 of the University of Granada, Spain.


\appendix
\section{Appendix}\label{ap}
In a simple linear model where the independent variables are the intercept  ($i_n$) and a dummy variable ($i_m$) with $m$ ones and being $n>m$, the matrix $\mathbf{X}=\left(
                                                                                                                                                                \begin{array}{cc}
                                                                                                                                                                  i_n & i_m \\
                                                                                                                                                                \end{array}
                                                                                                                                                              \right)
$ can be transformed into unit length obtaining:
\[\tilde{\mathbf{X}}=\left(
                                                                                                                                                                \begin{array}{cc}
                                                                                                                                                                  \sqrt{i_n}{n} & \sqrt{i_m}{m} \\
                                                                                                                                                                \end{array}
                                                                                                                                                              \right)\]
In this case, it is verified that:
\[\tilde{\mathbf{X}}^{t}\tilde{\mathbf{X}}=\left(
                        \begin{array}{cc}
                          1 & \frac{m}{\sqrt{nm}} \\
                          \frac{n}{\sqrt{nm}} & 1 \\
                        \end{array}
                      \right)=\left(
                        \begin{array}{cc}
                          1 & \sqrt{p}\\
                          \sqrt{p} & 1 \\
                        \end{array}
                      \right)\]
where $p$ is the proportion of ones existing in $i_m$. The eigenvalues of $\tilde{\mathbf{X}}^{t}\tilde{\mathbf{X}}$ are $1+\sqrt{p}$ and $1-\sqrt{p}$, respectively. Then, the condition number is calculated as $CN=\sqrt{\frac{1+\sqrt{p}}{1-\sqrt{p}}}$. Thus, when $p$ tends to one it is verified that the condition number tends to infinity.


%
%

\bibliographystyle{spmpsci}      
\bibliography{References}   

\end{document}